\documentclass[preprint,12pt]{elsarticle}
\usepackage{graphicx}
\usepackage{amssymb}
\usepackage{amsmath}
\usepackage{amsfonts}
\usepackage{slashed}
\usepackage[dvipsnames]{xcolor}
\usepackage{braket,bm}
\usepackage{multirow}
\usepackage{enumitem}
\usepackage[vcentermath]{youngtab}

\begin{document}

\begin{frontmatter} 

\title{Tribute to Toshimitsu Yamazaki (1934-2025): \\ Quest for 
Exotic Hadronic Matter} 
\author{Avraham Gal\corref{cor}~} 
\address{Racah Institute of Physics, The Hebrew University,
Jerusalem 9190400, Israel} 
\cortext[cor]{Avraham Gal,~~avragal@savion.huji.ac.il \\ 
Presented at the 21st Int'l Conf. Hadron Spectroscopy and 
Structure (HADRON 2025) \\ 
March 27-31, 2025, Toyonaka Campus, Osaka University, Japan \\ 
To be published as PoS(HADRON2025)263} 

\begin{abstract} 
In this talk I pay tribute to Toshimitsu Yamazaki who died earlier this year. 
Yamazaki's leading contributions to Hadronic Physics, in particular to 
Strangeness Nuclear Physics in Japan and elsewhere, are well known. Two 
of the five Recurring Themes of his research, as listed in the Japan Academy 
site, are highlighted here: (i) Discovery of deeply bound pionic-atom states, 
and (ii) Search for kaonic nuclei -- Kaonic Proton Matter (KPM). I conclude 
by reviewing briefly my own recent work, confirming Farrar's conjecture that 
a deeply bound $H$ dibaryon is not ruled out by the weak-decay observation 
of $\Lambda\Lambda$ hypernuclei. However, the relatively long lifetime of 
such a deeply bound $H$ is much too short to qualify it for a Dark-Matter 
candidate. 
\end{abstract} 

\end{frontmatter}   


\section{Toshimitsu Yamazaki: CV highlights} 
\label{sec:CV} 

\begin{itemize} 
\item Doctor of Science, University of Tokyo, 1964. 
\item Research Fellow, Berkeley \& Copenhagen, 1964-1967. 
\item Lecturer to Professor, Dept Phys, Univ Tokyo, 1967-1987. 
\item Director, Meson Science Lab, Univ Tokyo, 1978-1986. 
\item Professor \& Director, Inst Nucl Study, Univ Tokyo, 1986-1995; 
Emeritus since 1995, at RIKEN 2000-2021. 
\item Selected academic prizes: Matsunaga 1972, Nishina Memorial 1975, 
Japan Imperial \& Academy 1987, Fujiwara 1994; and several prestigious 
awards since 2000.
\end{itemize} 

\begin{figure}[!h]
\begin{center}
\includegraphics[width=0.4\textwidth]{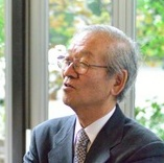}
\caption{Toshimitsu Yamazaki (1934-2025) was leading Strangeness Nuclear 
Physics worldwide, Japan in particular, for several decades. This is reflected 
in the rich \& imaginative experimental program in J-PARC, where many of his 
students have become leaders in their own right.} 
\label{Toshi_pdf}
\end{center}
\end{figure}

\section{Recurring Themes (from Japan Academy site)} 
\label{sec:themes} 

\begin{itemize} 
\item I.~~~Meson effects in nuclear magnetic moments. 
\item II.~~Muon spin rotation ($\mu$SR) relaxation. 
\item III.~Discovery of metastable antiprotonic helium. 
\item IV.~Discovery of deeply bound pionic-atom states. 
\item V.~~Search for kaonic nuclei; Kaonic Proton Matter (KPM). 
\end{itemize} 
Here I focus on the last two Recurring Themes, developed and matured 
throughout the last 40 years since I first met him at TRIUMF in 1985 
on my sabbatical stay there. 


\section{Deeply bound pionic atoms} 
\label{sec:pionic} 

\begin{figure}[!h] 
\begin{center} 
\includegraphics[width=0.48\textwidth,height=10cm]{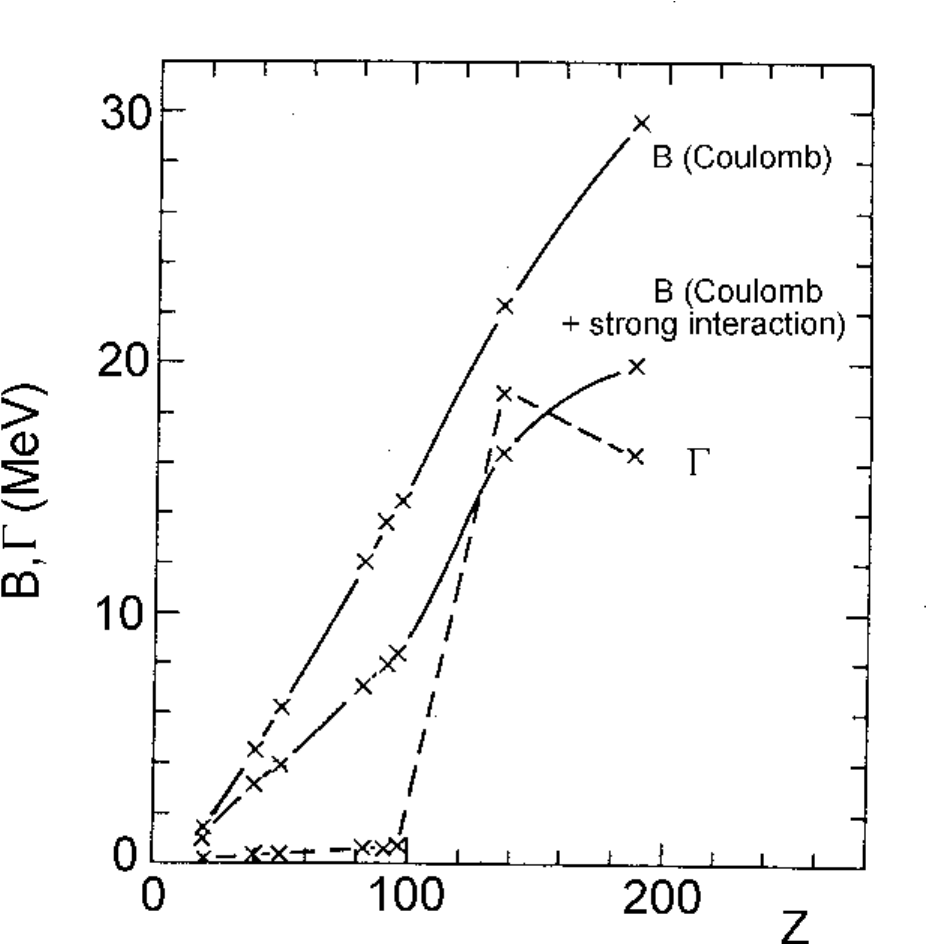} 
\includegraphics[width=0.48\textwidth]{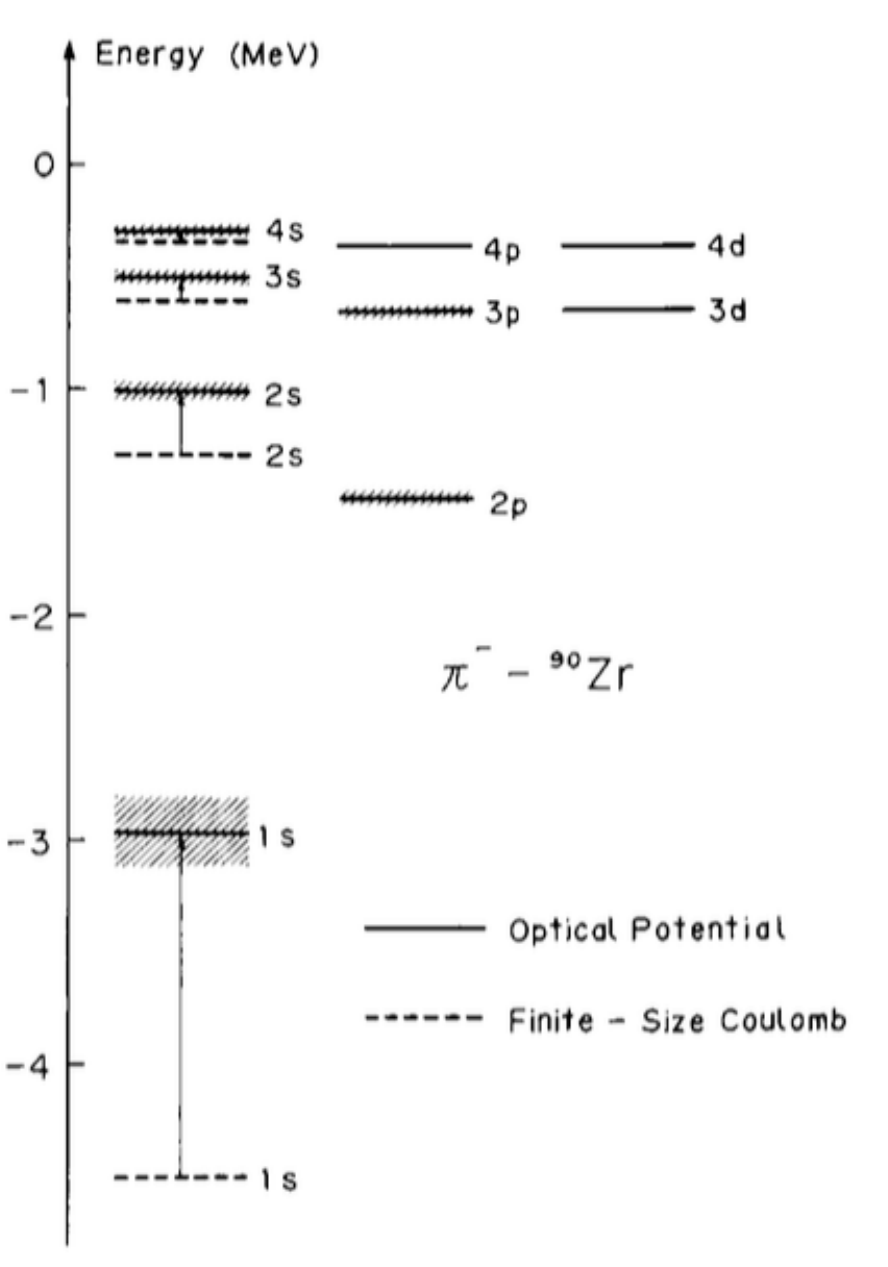} 
\caption{Left: binding energies (B) and widths ($\Gamma$) of 1s states in 
pionic atoms (Friedman \& Soff~\cite{FS85}). Right: $\pi^-$ energy levels in 
$^{90}$Zr with \& without Strong Interactions (Figure adapted from Toki \& 
Yamazaki~\cite{TY88}).} 
\label{fig:1s} 
\end{center} 
\end{figure}

Deeply bound pionic atoms refer to 1s and 2p states in heavy pionic atoms 
which cannot be reached in X-ray cascade because upper levels such as 3d 
are already broadened by Strong Interactions rendering the radiative yield 
exceedingly small. In 1985, Friedman and Soff~\cite{FS85} noted that deeply 
bound 1s pionic states are sufficiently narrow to make them well defined, 
see Fig.~\ref{fig:1s} (left) where the calculated 1s widths remain small 
up to the top end of the periodic table. The relative narrowness of the 1s 
states follows from the {\it repulsive} real part of the s-wave pion-nucleus 
potential that keeps the absorptive imaginary part off the nuclear volume.   

Three years later in 1988 Toki and Yamazaki~\cite{TY88}, unaware of the 
Friedman-Soff paper, realized that the width of 1s pionic states in heavy 
nuclei is considerably smaller than the 1s$\to$2p excitation energy, see 
Fig.~\ref{fig:1s} (right), so it made sense to look for strong-interaction 
reactions to populate such states. It took another eight years to apply 
a good `recoil-less' $^{208}$Pb(d,$^3$He) reaction at beam energy about 600 
MeV in GSI to create a deeply bound pion as close to rest as possible in 
$^{207}$Pb (Yamazaki, Hayano, et al.~\cite{YH96}). This was followed by 
experiments on $^{206}$Pb and on Sn isotopes, as reviewed by Kienle and 
Yamazaki~\cite{KY04}, Friedman and Gal~\cite{FG07} and Yamazaki, Hirenzaki, 
Hayano and Toki \cite{YHHT12}. More recently, both 1s and 2p deeply-bound 
pionic states in $^{121}$Sn were identified and studied at RIKEN~\cite{Sn18}.   


Pionic-atom states, deeply-bound as well as `normal' ones observed in atomic 
cascade, provide good evidence for a nuclear-medium renormalization of the 
isovector s-wave $\pi N$ scattering length $b_1$ in a form suggested by 
Weise~\cite{W00}: 
\begin{equation}
b_1(\rho)=b_1^{\rm free}(1-{\frac{\sigma_{\pi N}}{m_{\pi}^2 f_{\pi}^2}}
\rho)^{-1}, 
\label{eq:W00}
\end{equation} 
where $f_{\pi}=92.2$~MeV is the free-space pion decay constant and 
$\sigma_{\pi N}$ is the pion-nucleon sigma term. Fitting globally pionic 
atom energy shifts and widths across the periodic table, Friedman and Gal 
were able to extract the value $\sigma_{\pi N}=57\pm 7$~MeV~\cite{FG19}, 
in excellent agreement with a value 59.1$\pm$3.5~MeV derived using $\pi^-$H 
and $\pi^-$D atom data~\cite{Hofe15}, or 58$\pm$5~MeV derived using $\chi$EFT 
$\pi N$ scattering lengths~\cite{Hofe18}.

\section{Kaonic Proton Matter} 
\label{sec:KPM} 

Interpreting the $\Lambda$(1405) as a $K^-p$ quasibound state suggests that 
$K^-$ mesons are likely to bind strongly into nuclear clusters, the simplest 
of which is $K^-pp$. That a $J^{\pi}=0^-,I=\frac{1}{2}$ $K^-pp$ state might be 
bound by $\sim$10~MeV was suggested by Nogami already in 1963~\cite{Nogami63}. 
Yamazaki and Akaishi~\cite{YA02,AY02}, using a complex energy-independent 
${\bar K}N$ potential within a single-channel $K^-pp$ calculation, obtained 
binding energy $B_{K^-pp}\sim 50$~MeV and width $\Gamma_{K^-pp}\sim 60$~MeV. 
Subsequent ${\bar K}NN$--$\pi\Sigma N$ coupled-channel Faddeev calculations 
\cite{SGM07,SGMR07} confirmed this order of magnitude for $B$ but gave 
larger width values ($\sim$100~MeV). Using a chirally motivated ${\bar K}N$ 
interaction in such Faddeev calculations lowers $B_{K^-pp}$ to about 32~MeV 
and $\Gamma_{K^-pp}$ to about 50~MeV~\cite{RS14}, while few-body calculations 
using a single-channel effective chirally motivated ${\bar K}N$ interaction 
find even smaller values of binding energies and widths~\cite{DHW08,BGL12}. 
Experimentally, following several dubious claims by several experiments, 
J-PARC E15 reported a statistically reliable $K^-pp$ signal~\cite{Yam20} with 
$B_{K^-pp}=42\pm 3^{+3}_{-4}$~MeV, $\Gamma_{K^-pp}=100\pm 7^{+19}_{-9}$~MeV. 
Part of this large width comes from $K^- NN$ absorption processes that are 
not accounted for in most calculations. 

Calculations of $\bar K$ nuclear clusters up to six nucleons~\cite{Ohnishi17} 
demonstrate that replacing one nucleon by a $\bar K$ meson increases 
substantially the overall binding energy. This suggests to look for a maximal 
increase by considering aggregates of bound $\Lambda^{\ast}$ hyperons, 
so-called Kaonic Proton Matter (KPM) by Akaishi and Yamazaki~\cite{AY17} who 
argued that it would also provide an absolutely stable form of matter for less 
than ten $\Lambda^{\ast}$ hyperons. This exciting proposal was questioned by 
the Jerusalem-Prague collaboration~\cite{Jarka18,Jarka19,Jarka20} within 
a Relativistic Mean Field (RMF) calculational scheme, as follows. 

\begin{figure}[htb] 
\begin{center} 
\includegraphics[width=0.48\textwidth]{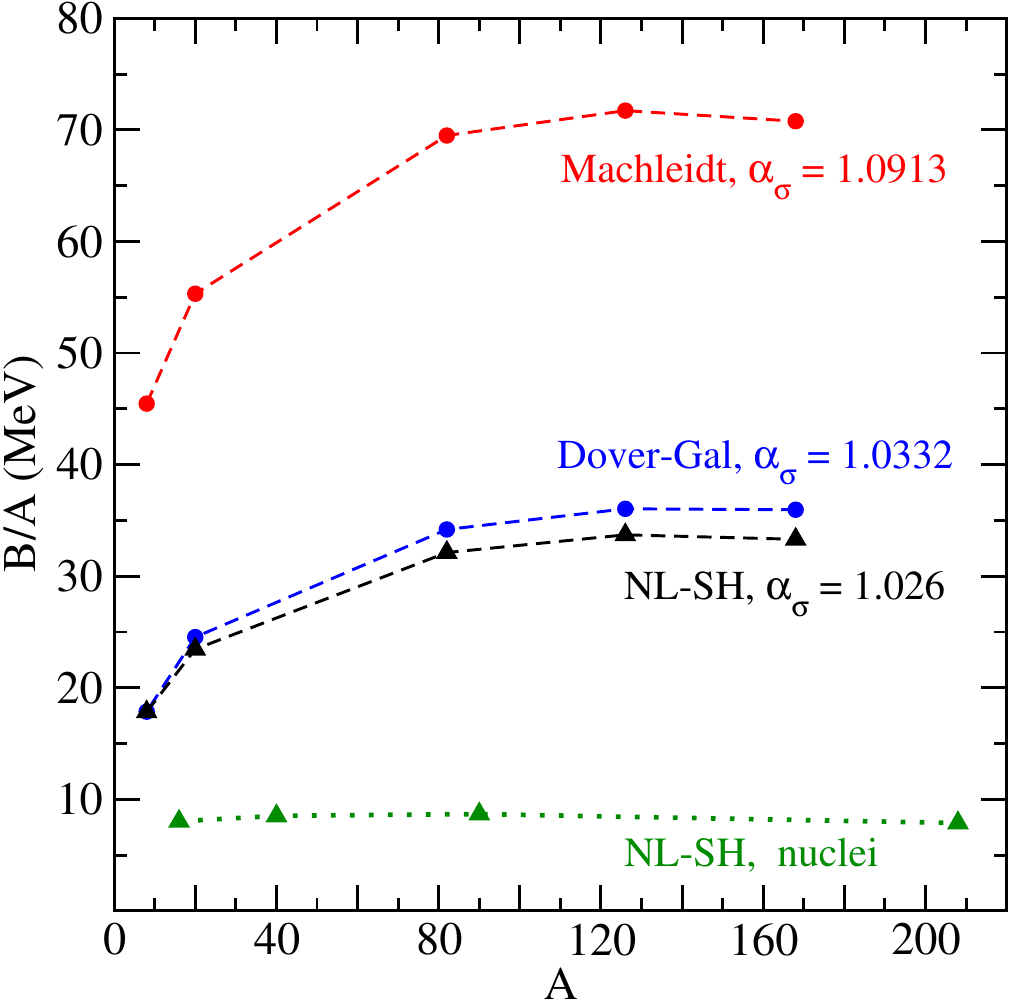} 
\includegraphics[width=0.48\textwidth]{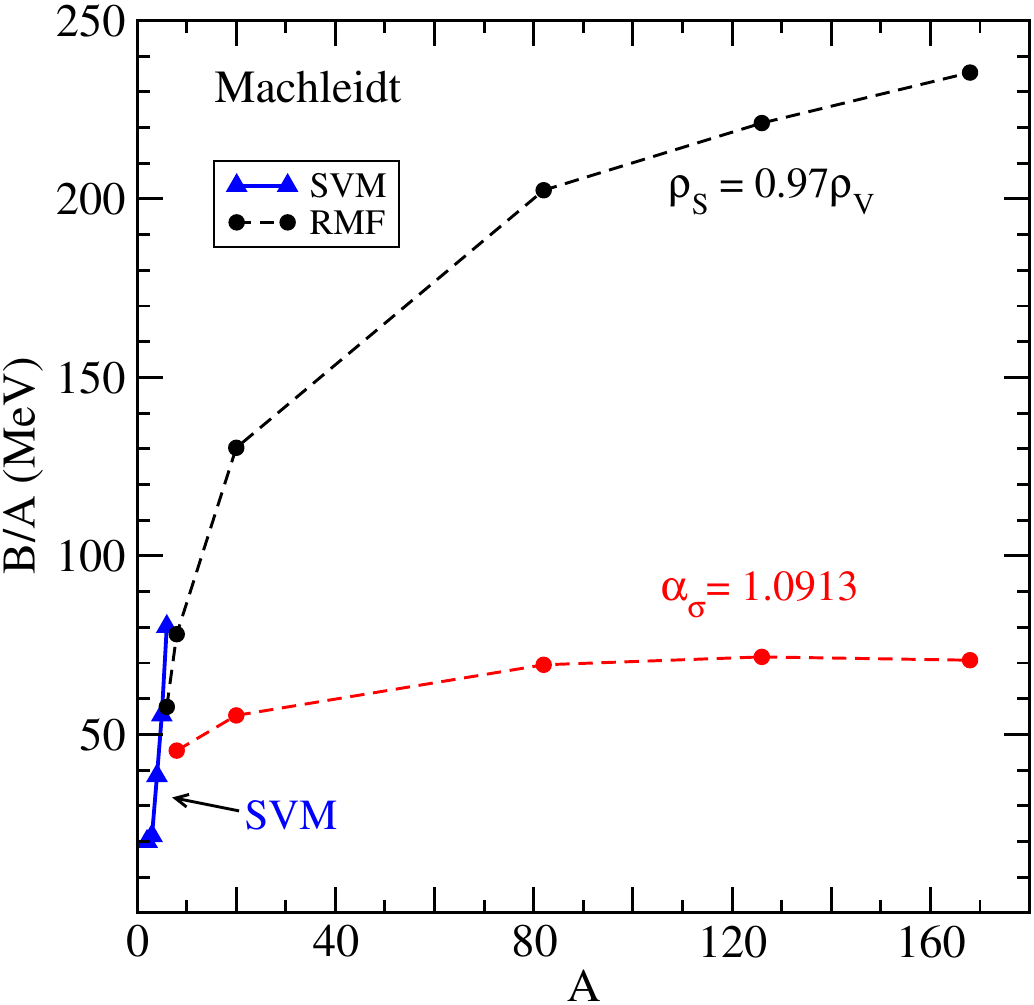} 
\caption{Left: binding energies per particle ($B/A$) of $\Lambda^{\ast}$ 
nuclei as function of mass number $A$, calculated in two RMF versions, 
saturating at $B/A \approx 70$ (Machleidt) or 35~MeV (Dover-Gal). $B/A$ values 
calculated for atomic nuclei are shown for comparison. Right: $B/A$ for 
$\Lambda^{\ast}$ nuclei from the left-hand side calculation (both in red), 
compared with a similar calculation (in black) using $\rho_s=0.97\rho_v$. 
$B/A$ values calculated in few-$\Lambda^{\ast}$ systems, marked SVM, are shown 
for comparison. Figure adapted from Fig.~2, Ref.~\cite{Jarka20}.} 
\label{fig:RMF} 
\end{center} 
\end{figure} 

The $\Lambda^{\ast}\Lambda^{\ast}$ interaction strength used as input to these 
RMF calculations of $\Lambda^{\ast}$ nuclei was constrained by a two-body 
binding energy value $B(\Lambda^{\ast}\Lambda^{\ast})\approx 40$~MeV, deduced 
from $B(K^-K^-pp)=93$~MeV following an earlier work by Maeda, Akaishi and 
Yamazaki~\cite{MAY13}. It was found then, as shown on the left-hand side of 
Fig.~\ref{fig:RMF}, that the binding energy per $\Lambda^{\ast}$ saturates 
at values well below 100 MeV for mass number $A\geq 120$, implying that 
$\Lambda^{\ast}$ matter is highly unstable against strong-interaction decay 
to $\Lambda$ and $\Sigma$ aggregates. Recall that about 300 MeV is required 
to reduce the $\Lambda^{\ast}(1405)$ mass in the medium below that of the 
lightest hyperon $\Lambda(1116)$. It is worth noting that RMF calculations for 
multi $\bar K$-N-$\Lambda$ hadronic systems reported earlier by Gazda et 
al.~\cite{GFGM07,GFGM08,GFGM09} reached similar conclusions, although in the 
context of ruling out kaon condensation. 


The saturation of $B/A$ observed on the left-hand side of Fig.~\ref{fig:RMF} 
follows from the decrease of the scalar density $\rho_s$ associated with 
the attractive $\sigma$ field with respect to the conserved vector density 
$\rho_v$ associated with the repulsive $\omega$ field: 
$\rho_s=(M^{\ast}/E^{\ast})\rho_v$ < $\rho_v$, 
where $M^{\ast}=M-g_{\sigma B}{\bar\sigma}$ is the baryon-$B$ effective mass. 
Thus, Lorentz invariance implies that the scalar-field attraction is reduced 
as density increases. The upper curve on the right-hand side of 
Fig.~\ref{fig:RMF} demonstrates then how a {\it fixed} ratio 
$(\rho_s/\rho_v)=0.97$, corresponding to $^{16}$O calculation, leads to 
non-saturating $B/A$ values in contrast to the saturated values obtained in 
the properly calculated RMF lower curve. It is worth noting that the central 
density of $\Lambda^{\ast}$ matter is found to saturate as well, at roughly 
twice nuclear-matter density.

\section{A deeply bound $H$ dibaryon?} 
\label{sec:H} 

Lattice-QCD (LQCD) calculations 
suggest two strong-interaction stable hexaquarks. Both are $J^{\pi}$=0$^+$ 
near-threshold $s$-wave dibaryons with zero spin and isospin: (i) a maximally 
strange $ssssss$ hexaquark classified as $\Omega\Omega$ dibaryon member of the 
SU(3) flavor ${\bf 28}_f$ multiplet, and (ii) a doubly strange $S=-2$ $uuddss$ 
hexaquark, a ${\bf 1}_f$ $H$ dibaryon. Whereas the LQCD calculation of 
$\Omega\Omega$ reached $m_{\pi}$ values close to the physical pion 
mass~\cite{LQCD18}, $H$ dibaryon LQCD calculations have been limited 
to values of $m_{\pi}\sim 400$~MeV and higher (NPLQCD~\cite{LQCD11a}, 
HALQCD~\cite{LQCD11b}) while following SU(3)$_f$ symmetry, where 
\begin{equation} 
H = -\sqrt{\frac{1}{8}}\Lambda\Lambda +\sqrt{\frac{3}{8}}\Sigma\Sigma 
+\sqrt{\frac{4}{8}}N\Xi. 
\label{eq:H} 
\end{equation} 
A recent calculation of this type~\cite{LQCD21} finds the $H$ dibaryon
bound just by 4.6$\pm$1.3~MeV with respect to the $\Lambda\Lambda$ threshold.
However, chiral extrapolation to physical quark-mass values and thereby also
to $m_{\pi}\approx 0$~\cite{LQCD11c} suggests that the $H$ dibaryon becomes
{\it unbound} by 13$\pm$14~MeV. Thus, a slightly bound $\bf{1}_f$ $H$ dibaryon
is likely to become unbound with respect to the $\Lambda\Lambda$ threshold in
the SU(3)$_f$-broken physical world, lying possibly a few MeV below the $N\Xi$
threshold~\cite{LQCD12,EFT12}.

The $H$ dibaryon was predicted in 1977 by Jaffe~\cite{Jaffe77} to lie about 
80~MeV below the $\Lambda\Lambda$ threshold. Dedicated experimental searches, 
beginning as soon as 1978 with a $pp\to K^+K^-X$ reaction~\cite{AGS78} at BNL, 
have failed to observe $S=-2$ dibaryon signal over a wide range of dibaryon 
masses below 2$m_{\Lambda}$~\cite{Belle13,ALICE16,BABAR19}. 
A particularly simple argument questioning its existence was given by 
Dalitz et al.~\cite{Dalitz89}. It involves the lightest known $\Lambda\Lambda$ 
hypernucleus~\cite{GHM16} where a $\Lambda\Lambda$ pair is bound to $^4$He by 
6.91$\pm$0.17~MeV. If $H$ existed deeper than about 7~MeV below the 
$\Lambda\Lambda$ threshold, ${_{\Lambda\Lambda}^{~~6}}$He could decay 
{\it strongly}, 
\begin{equation} 
{_{\Lambda\Lambda}^{~~6}}{\rm He} \to {^4{\rm He}} + H, 
\label{eq:LL6He} 
\end{equation} 
considerably faster than the $\Delta S=1$ {\it weak-interaction} decay 
by which it has been observed and uniquely identified~\cite{Ahn13}: 
\begin{equation} 
{_{\Lambda\Lambda}^{~~6}}{\rm He} \to {_{\Lambda}^5}{\rm He}+p+\pi^-. 
\label{eq:L5He} 
\end{equation} 

Treating ${_{\Lambda\Lambda}^{~~6}}$He in a $\Lambda-\Lambda-{^4}$He 3-body 
model, I confirmed~\cite{Gal24} that the ${_{\Lambda\Lambda}^{~~6}}{\rm He}\to 
H+{^4{\rm He}}$ strong-interaction lifetime is correlated strongly with the 
$H$ mass ($m_H$), increasing upon decreasing $m_H$ such that it exceeds the 
hypernuclear $\Delta S=1$ weak-decay lifetime of order 10$^{-10}$~s for $H$ 
masses $m_H < (m_{\Lambda}+m_n)$. 
Therefore, and as argued by Farrar~\cite{Farrar03}, hypernuclear physics by 
itself does not rule out the occurrence of an $H$-like dibaryon in this mass 
range. This conclusion is robust against varying the inner structure of the 
$H$ hexaquark, such as its size, within reasonable limits. 

\begin{figure}[!h] 
\begin{center} 
\includegraphics[width=0.5\textwidth]{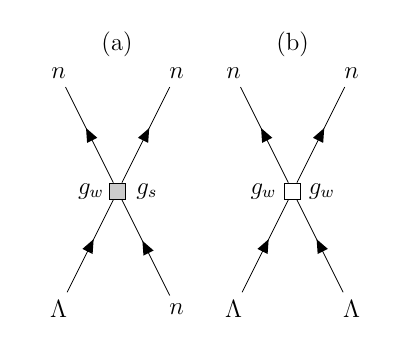} 
\caption{$^1S_0\to {{^1}S_0}$ LO EFT $\Delta S\neq 0$ weak-interaction 
diagrams: (a) $\Delta S=1$ $\Lambda n\to nn$, (b) $\Delta S=2$ 
$\Lambda\Lambda\to nn$.} 
\label{fig:nmwd} 
\end{center} 
\end{figure} 

Having realized that a deeply bound $H$ dibaryon lying below $m_{\Lambda}+m_n$
is not in conflict with the weak-decay lifetime scale $\tau_{\Lambda}\sim
10^{-10}$~s of all observed $\Lambda\Lambda$ hypernuclei, one proceeds to 
estimate the leading $\Delta S=2$ $H\to nn$ weak-interaction decay rate, where 
$H$ is represented by its deeply bound $\Lambda\Lambda$ component. 
Although $\Delta S=2$ $\Lambda\Lambda\to nn$ transitions are not constrained 
directly by experiment, they are related to $\Delta S=1$ $\Lambda n\to nn$ 
transitions which are constrained by ample lifetime data in $\Lambda$ 
hypernuclei~\cite{GHM16}. It was found useful to follow Effective-Field-Theory 
(EFT) approach~\cite{PBH04} with Leading-Order (LO) Low-Energy-Constants 
(LECs) denoted in Fig.~\ref{fig:nmwd} schematically by weak-interaction and 
strong-interaction coupling constants $g_w$ and $g_s$, respectively. For an 
order-of-magnitude estimate one takes $g_s=1$ and $g_w=G_Fm_{\pi}^2=2.21\times 
10^{-7}$, where $G_F$ is the Fermi weak-interaction constant.     

Constrained by $\Lambda$ hypernuclear $\Delta S=1$ nonmesonic weak-interaction 
decay rates within LO EFT approach, a realistic calculation of 
the $\Delta S=2$ $H\to nn$ weak decay for $H$ mass satisfying 
$2m_n \lesssim m_H < (m_n+m_{\Lambda})$ results then in $H$ lifetimes of order 
$10^5$~s, 10 orders of magnitude shorter than the order of 10$^8$ yr claimed 
in 2004 by Farrar~\cite{FZ04}. Our result is in rough agreement with Donoghue, 
Golowich, Holstein~\cite{DGH86} who followed a rather different high-energy 
physics methodology. Hence, such a deeply bound $H$ dibaryon would be far 
from qualifying for a Dark-Matter candidate. This conclusion holds also to 
a lower-mass $H$, below $2m_n$, where two neutrons could decay comfortably 
to $H$ by a $\Delta S=2$ weak decay of $^{16}$O, say in $^{16}{\rm O}\to H + 
{^{14}{\rm O}}$, thereby defying the known $^{16}$O nuclear stability limit.

\section*{Acknowledgements}  
I would like to thank the Organizers of HADRON 2025 for inviting me to 
give this Tribute talk. Thanks are also due to Eliahu Friedman, Jaroslava 
{\'O}bertov{\'a} (Hrt{\'a}nkov{\'a}) and Hiroshi Toki for providing me with 
adapted versions of published figures.

\end{document}